\documentclass[final,5p,times,twocolumn]{elsarticle}

\usepackage{graphicx}
\graphicspath{{./pdf/}{./jpeg/}{./images/}}
   \DeclareGraphicsExtensions{.pdf,.jpeg,.jpg,.png}
\usepackage{amssymb}
\usepackage[nodots,nocompress]{numcompress}

\usepackage{textcomp}

\usepackage{upgreek}

\usepackage{url}

\usepackage{amsmath}

\usepackage[T1]{fontenc}

\usepackage{array}

\biboptions{numbers}

\journal{Digital Forensics Research Workshop EU 2014}

\begin{document}

\begin{frontmatter}

\title{BitTorrent Sync: First Impressions and Digital Forensic Implications}

\author{Jason Farina}
\ead{jason.farina@ucdconnect.ie}
\author{Mark Scanlon}
\ead{mark.scanlon@ucd.ie}
\author{M-Tahar Kechadi}
\ead{tahar.kechadi@ucd.ie}
\address{UCD School of Computer Science and Informatics,\\
		University College Dublin, Dublin 4, Ireland.}

%\author{Removed for blind review.}
%\ead{Removed for blind review.}
%\address{Removed for blind review.}

\begin{abstract}
With professional and home Internet users becoming increasingly concerned with data protection and privacy, the privacy afforded by popular cloud file synchronisation services, such as Dropbox, OneDrive and Google Drive, is coming under scrutiny in the press. A number of these services have recently been reported as sharing information with governmental security agencies without warrants. BitTorrent Sync is seen as an alternative by many and has gathered over two million users by December 2013 (doubling since the previous month). The service is completely decentralised, offers much of the same synchronisation functionality of cloud powered services and utilises encryption for data transmission (and optionally for remote storage). The importance of understanding BitTorrent Sync and its resulting digital investigative implications for law enforcement and forensic investigators will be paramount to future investigations. This paper outlines the client application, its detected network traffic and identifies artefacts that may be of value as evidence for future digital investigations.
\end{abstract}


\end{frontmatter}

% \linenumbers

%%%%%%%%%%%%%%%%%%%%%%%%%%%%%%%%%%%%%%%%%%%%%%%%%%%%%%%%%%%%%%%%%%%%%%
\section{Introduction}
\label{introduction}
%%%%%%%%%%%%%%%%%%%%%%%%%%%%%%%%%%%%%%%%%%%%%%%%%%%%%%%%%%%%%%%%%%%%%%

With home user bandwidth rising and the growth in professional and non-professional computer power, the volume of data created by each individual computer user is constantly growing. For mobile users, access to this data has long been an issue. With greater connectivity and greater availability of access to the Internet the concepts of ``high availability'', ``off-site backup'' and ``resilient storage'' have moved away from the domain solely inhabited by large corporations and has started to become increasingly popular with computer users and everyday data consumers. Applications such as Evernote and Dropbox leverage the decreasing cost of hard disk storage seen in Storage as a Service (SaaS) providers, e.g., Amazon S3, to provide data storage on the cloud to home users and businesses alike. The main advantage of services such as Dropbox, Google Drive, Microsoft Skydive and Apple iCloud to the end user is that their data is stored in a virtual extension of their local machine with no direct user interaction required after installation. It is also backed up by a full distributed data-centre architecture that would be completely outside the financial reach of the average consumer. Their data is available anywhere with Internet access and is usually machine agnostic so the same data can be accessed on multiple devices without any need to re-format partitions or wasting space by creating multiple copies of the same file for each device. Some services such as Dropbox, also have offline client applications that allow for synchronisation of data to a local folder for offline access.

Each of the aforementioned services can be categorised as cloud synchronisation services. This means that while the data is synchronised between user machines, a copy of the data is also stored remotely in the cloud. In recent headline news, much of this data is freely available to governmental agencies without the need of a warrant or even just cause. As a result, BitTorrent Sync (also referred to as BTSync, BitSync or BSync), which provides much of the same functionality without the cloud storage aspect is seen by many as a real alternative. The service has numerous desirable attributes for any Internet user \cite{bitsync}:

\begin{itemize}
\item Compatibility and Availability -- Clients are built for most common desktop and mobile operating systems, e.g., Windows, Mac OS, Linux, BSD, Android and iOS.
\item Synchronisation Options -- Users can choose whether to sync their content over a local network or over the Internet to remote machines.
\item No Limitations or Cost -- Most cloud synchronisation services provide a free tier offering a small amount of storage and subsequently charge when the user outgrows the available space. BTSync eliminates these limitations and costs. The only limitation to the volume of storage and speed of the service is down to the limitations of the synchronised users machines.
\item Automated Backup -- Like most competing products, once the initial install and configuration is complete, the data contained within specified folders is automatically synchronised between machines.
\item Decentralised Technology -- All data transmission and synchronisation takes place solely in a Peer-to-Peer (P2P) fashion, based on the BitTorrent file sharing protocol.
\item Encrypted Data Transmission -- While synchronising data between computers on different LANs (the option exists to apply encryption for internal LAN transmission), the data is encrypted using RSA encryption. Under the BTSync API \cite{bitsyncapi}, developers can also enable remote file storage encryption. This could result in users storing their data on untrusted remote locations for the purposes of redundancy and secure remote backup.
\item Proprietary Technology -- The precise protocol and operation of the technology is not openly documented by the developer resulting in an element of perceived ``security through obscurity''. Of course, this requires a significant degree trust on behalf of users that the developers' security has been implemented and tested correctly.
\end{itemize}

As a result of the above, the BTSync application has become a very popular choice for file replication and synchronisation. The technology had grown to over one million users by November 2013 and doubled to over two million users by December 2013 \cite{bitsyncstats}. The service will be of interest to both law enforcement and digital forensics investigators in future investigations. Like any other file distribution technology, this interest may be centred around recovering evidence of the data itself, of the modification of the data or of where the data is synchronised to. While the legitimate usage of the system, e.g., backup and synchronisation, teamwork, data transfer between systems, etc., may be of interest to an investigation, the technology may also be a desirable one for a number of potential crimes including industrial espionage, copyright infringement, sharing of child exploitation material, malicious software distribution, etc.

\subsection{Contribution of this work}
\label{contribution}
This contribution of this work includes a forensic analysis of the BTSync client application, its behaviour, artefacts created during installation and use, and remnants left behind after uninstallation. An analysis of the sequence of network traffic and file I/O interactions used as part of the synchronisation process are also provided. This information should prove useful to digital forensic investigators when BTSync is found to be installed on a machine under investigation. Gaining an understanding of how BTSync operates could aid in directing the focus of a digital investigation to additional remote machines where any pertinent data is replicated. Depending on the crime under investigation, these remote machines may be owned and operated by a single suspect or by a group sharing a common goal. While an initial analysis of the network protocol and its operation is included below, comprehensive network analysis is beyond the scope of this paper.

%%%%%%%%%%%%%%%%%%%%%%%%%%%%%%%%%%%%%%%%%%%%%%%%%%%%%%%%%%%%%%%%%%%%%%
\section{Background}
\label{background}
%%%%%%%%%%%%%%%%%%%%%%%%%%%%%%%%%%%%%%%%%%%%%%%%%%%%%%%%%%%%%%%%%%%%%%

In order to understand how BTSync operates, its important to first understand the technologies its based upon and how a number of similar technologies operate. This section provides some of the required background information.

\subsection{BitTorrent File Sharing Protocol}
\label{bittorrent}

The BitTorrent protocol was designed with the aim of facilitating one-to-many and many-to-many file transfers as efficiently as possible. The protocol is described in BitTorrent Enhancement Proposal (BEP) No. 3 \cite{cohen2008bittorrent}. The main strength of the protocol is the usage of file parts, each of which can be manipulated and managed separately. While one part of a file downloads, another, already downloaded part can be uploaded to a different peer. In this way, peers can start trading parts even before they have downloaded the entire file themselves. This has the benefit of not only speeding up distribution as each peer can find useful information on a broad range of potential peers but it also helps alleviate the issues of ``churn'' \cite{Stutzbach:2006:UCP:1177080.1177105} and ``free riding'' \cite{4797943} experienced with older protocols such as Gnutella and eDonkey. Data leeching is where a user downloads an entire file in one go and then removes the share to avoid uploading. Data churn is the natural expansion and retraction of the network horizon as peers leave and join the "swarm" freely resulting in a large variance in the availability of full versions of a file being available from individual sources.

The overall BitTorrent network can be seen as being sub-divided into BitTorrent ``swarms''. Each swarm consists of a collection of peers involved in the sharing of the same file. The central commonality of a swarm is a unique identifier created from a SHA-1 hash of the file(s) references in the metadata. A peer can be a member of multiple swarms as multiple files are uploaded and downloaded simultaneously. In order to initiate download of content from a particular swarm the user must first download a metadata \texttt{.torrent} file (or corresponding magnet URI) from an indexing website. The BitTorrent client application running on the users machine then interprets the metadata and uses it to locate other peers actively participating in that swarm using one or more of the following methods \cite{scanlon2010week}:

\begin{enumerate}
\item Tracker Server -- Tracker servers are Internet accessible servers that maintain a list of \texttt{seeders} (those peers with 100\% of a file available and as such are only uploading data) and \texttt{leechers} (peers that are beginning the process or are in the middle of the process of downloading information from the swarm) \cite{cohen2003incentives}. While the data transfer is in progress, the client application will periodically report to the tracker to update its status and to update its list of active peers.
\item Distributed Hash Table (DHT) -- While the original BitTorrent protocol was designed with central repositories of peers stored on servers, clients were developed such as Vuze and $\upmu$Torrent that also stored a list of active clients on the local machine. This common DHT allows peers to identify peers through requesting information from other BitTorrent clients without the requirement for a central server (these clients serving information from the DHT are likely not involved in the requested swarm). Each peer record in the DHT is associated with the swarms in which it is actively participating. The Mainline DHT, as outlined in BEP No. 5 \cite{cohen2008bittorrent}, that is used by BitTorrent and BTSync is based on the Kademlia protocol and allows for completely decentralised discovery of peers associated with sharing a particular piece of content (identified by the SHA-1 hash of the content).
\item Peer Exchange (PEX) -- Originally, the BitTorrent protocol did not allow for any direct communication between peers beyond the transmission of data, but various extensions of the protocol have resulted in the removal of this restriction. As DHT participation became commonly supported in the major BitTorrent clients, peers started to exchange the local peer caches. Peer Exchange is a BEP outlined a method for when two peers are communicating (sharing the data referenced by a torrent file), a subset of their respective peer lists are shared back and forth as part of the communication. Coupled with DHT, PEX removes a potential vulnerability from the BitTorrent network by allowing for fully distributed bootstrapping, tracking and peer discovery. 
\end{enumerate}

Any metadata or network control requests/responses are transmitted using ``bencoding'', as explained in BEP No. 3 \cite{cohen2008bittorrent}. Bencoded data consists of dictionaries and lists consisting of \texttt{key:value} pairs. Each key name and corresponding value is prepended by the length (in bytes) followed by a colon. For example the \texttt{get\_peers} request message can be bencoded as \texttt{1:m9:get\_peers} (with the `m' representing the key name ``message'').

\subsection{BitTorrent Sync}
\label{btsync}

BTSync is a file replication utility created by BitTorrent Inc. and released as a private alpha in April 2013\cite{bitsync}. It is not a cloud backup solution, nor necessarily intended as any form of offsite storage. Any data transferred using BTSync resides in whole files on at least one of the synchronised devices. This makes the detection of data much simpler for digital forensic purposes as there is no distributed file system, redundant data block algorithms or need to contact a cloud storage provider to get a list of all traffic to or from a container using discovered credentials. The investigation remains an examination of the local suspect machine. However, because BTSync optionally uses a DHT to transfer data there is also no central authority to manage authentication or log data access attempts. A suspect file found on a system may have been downloaded from one or many sources and may have been uploaded to many recipients. 

While the paid cloud synchronisation services offer up to 1TB of storage (Amazon S3 paid storage plan) the free versions which are much more popular with home users cap at approximately 10GB. The BTSync storage is limited only by the size of the folder being set as a share (most likely limited by the available disk space). Unless the system under investigation is the creator of the shared folder, it is possible that any files contained therein may have been downloaded without the user's prior knowledge of the folder's contents. The BTSync application does not feature a built in content preview utitily. As a result, it blindly and completely synchronises all content within the shared folder without any file selection process available to the user.

%%%%%%%%%%%%%%%%%%%%%%%%%%%%%%%%%%%%%%%%%%%%%%%%%%%%%%%%%%%%%%%%%%%%%%
\section{Related Work}
\label{related}
%%%%%%%%%%%%%%%%%%%%%%%%%%%%%%%%%%%%%%%%%%%%%%%%%%%%%%%%%%%%%%%%%%%%%%

\begin{figure}
\centering
\includegraphics[width=0.5\textwidth]{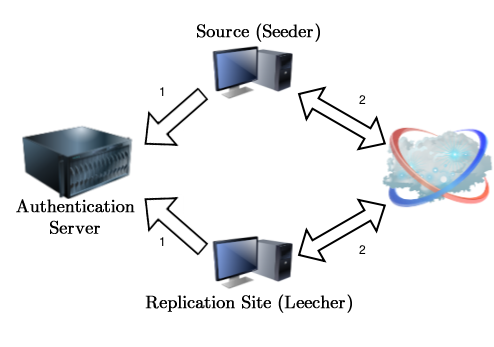}
\caption{Operation of Cloud File Synchronisation Services}
\label{figcloud}
\end{figure}

At the time of publication, there are no academic publications focussing on BTSync. However, due to BTSync sharing a number of attributes and functionalities with cloud synchronisation services, e.g., Dropbox, Google Drive, etc., and it is largely based on the BitTorrent protocol, there are a number of relevant related topics of interest. This section outlines a number of related case studies and investigative techniques for these shared technologies. While the specific attributes of a number of popular cloud synchronisation services are outlined below, there is a common generalised architecture employed by these services. There are two main stages to this synchronisation process, as shown in Figure~\ref{figcloud}: 

\begin{itemize}
\item Stage 1 -- The local client with the source file (the seeder in P2P terms) and the remote replication target (leecher) both contact the server of authority belonging to the service being used to confirm their credentials.
\item Stage 2 -- Both seeder and leecher contact the remote storage location, usually cloud based for high availability. The Seeder uploads a full copy of each file to be replicated and the leecher downloads a full version of the files it finds in the cloud storage container.
\end{itemize}
At no point in the process do the clients have to talk directly to one another. An important feature of these services is the fact that there is a full copy of the data being stored on a remote third party server outside the control of either client.

\subsection{Forensic Analysis of Cloud Synchronisation Clients}
\label{clientside}

Forensic investigation of these utilities can be challenging, as presented by Chung et al. in their 2012 paper \cite{Chung201281}. Unless local synchronisation is completely up to date, the full picture of the data may reside across temporary files, volatile storage (such as the system's RAM) and across multiple data-centres of the service provider's cloud storage facilities. Any digital forensic examination of these systems must pay particular attention to the method of access, e.g., usually the Internet browser connecting to the service provider's access page. This temporary access serves to highlight the importance of live forensic techniques when investigating a suspect machine. Cutting power to the suspect machine may not only lose access to any currently opened documents, but would also lose any currently stored passwords or other authentication tokens that are stored in RAM. Chung et al. describe three main forms of online storage in use by consumers:

\begin{enumerate}
\item Data Storage for Large Data -- Examples would include the services provided by Amazon S3, Dropbox, Google drive, etc.
\item Online Only Office Applications -- This includes services whereby an entire productivity suite of tools is accessed in a completely self contained online environment, e.g., Google Docs, Office 365 or Sage Online.
\item Personal Data -- Examples would include Evernote, which allows users to save notes to a central store, and Spotify, which allows playlists to be stored in the cloud when users build their online music catalogue.
\end{enumerate}

\subsection{Cloud File Synchronisation Services}
\label{cloudforensics}

In various complimentary papers on data remnants \cite{Quick2013266, Quick2013, quick2013digital}, Quick et. al offers an additional approach to forensics when dealing with cloud storage investigation. This involves accessing using the full client application whether or not it has been tampered with by the end user, e.g., perhaps an anti-forensics attempt was made to hide data by uninstalling the application and deleting the synchronised folders. Each of the applications examined stored their authentication credentials on the local system while the client was actively connected to the service, again highlighting the importance of live forensic recovery techniques. It should be noted that while Dropbox and Microsoft OneDrive (formally SkyDrive) appear to be very similar utilities, there are distinct differences in the way they are intended to be used. Dropbox (when used with the client application) creates a local folder that synchronises any contents stored in it with an online duplicate of that folder. By default, Dropbox gives 2GB of storage for free with an option to buy additional storage. OneDrive on the other hand is intended as a predominantly online storage facility with an option to synchronise a copy of the files to a client machine folder. However, this is not the default behaviour and has to be specifically enabled if used as part of the Windows 8.1 operating system. For non-Windows 8 based computers, the user is required to download and install the OneDrive desktop application to enable file synchronisation across devices.

Many Cloud storage utilities provide a method of synchronisation of files which involves some form of periodic checking to determine if changes have been made to any version being viewed locally or to compare offline copies with their online counterparts as soon as communication can be re-established (network connectivity re-enabled or the application or service restarted). For Dropbox, Drago et al. \cite{Drago:2012:IDU:2398776.2398827} identified two sets of servers, the control servers owned and operated by Dropbox themselves and the storage management and cloud storage servers hosted by Amazon's EC2 and S3 services. This identification is also verified by Wang et al. \cite{wang2012impact}. 

%%%%%%%%%%%%%%%%%%%%%%%%%%%%%%%%%%%%%%%%%%%%%%%%%%%%%%%%%%%%%%%%%%%%%%
\section{BTSync Application \& Protocol Analysis}
\label{analysis}
%%%%%%%%%%%%%%%%%%%%%%%%%%%%%%%%%%%%%%%%%%%%%%%%%%%%%%%%%%%%%%%%%%%%%%
\begin{table}[!h]
    \begin{tabular}{l|l|l|l}
      \textbf{Name} & \textbf{Host 1:}                 &  \textbf{Guest 1:}               \\ \hline
    OS     & Windows 7 PC ( 64 bit)  & Windows XP SP3            \\ \hline
    Ram    & 8GB ram                 & 512mb RAM                 \\ \hline
    ~      & Vmware Workstation 8    & Bridged network adapter   \\ \hline
    ~      & ~                       & ~                         \\
      \textbf{Name} & \textbf{Host 2:}                & \textbf{Guest 2:}                  \\ \hline
    OS     & Linux Debian laptop    & Widows XP SP3             \\ \hline
    Ram    & 4GB ram                & 512mb RAM                 \\ \hline
    ~      & VirtualBox 4.2         &   Bridged network Adapter \\
    \end{tabular}
\caption{Hardware Used in the Analysis of the BitTorrent Sync Application}
\label{tab:hardware}
\end{table}

\begin{table*}[t!]
\makebox[\linewidth]{
    \begin{tabular}{|l|l|}
	\hline
      	\textbf{File} & \textbf{Purpose}  \\ \hline
\$Volume\$\textbackslash Program Files\textbackslash BitTorrent Sync\textbackslash BTSync.exe &  Main Executable \\ \hline
\$Volume\$\textbackslash Documents and Settings\textbackslash [User]\textbackslash Application Data\textbackslash Microsoft\textbackslash Crypto\textbackslash<user SID> & Private Key\\ \hline
\$Volume\$\textbackslash Documents and Settings\textbackslash [User]\textbackslash Application Data\textbackslash Bittorrent Sync & Application folder  \\ \hline
\$Volume\$\textbackslash Documents and Settings\textbackslash [User]\textbackslash Application Data\textbackslash Bittorrent Sync\textbackslash settings.dat & Configuration Settings \\ \hline
\$Volume\$\textbackslash Documents and Settings\textbackslash [User]\textbackslash Application Data\textbackslash Bittorrent Sync\textbackslash sync.log & Log of Synchronisation Activity \\ \hline
\$Volume\$\textbackslash Documents and Settings\textbackslash [User]\textbackslash Application Data\textbackslash Bittorrent Sync\textbackslash sync.lng &  Language File \\ \hline
\$Volume\$\textbackslash Documents and Settings\textbackslash All Users\textbackslash Desktop\textbackslash BitTorrent Sync.lnk & Application Shortcut\\ \hline
\$Volume\$\textbackslash Documents and Settings\textbackslash All Users\textbackslash Start Menu\textbackslash BitTorrent Sync.lnk & Application Shortcut\\ \hline
\$Volume\$\textbackslash Documents and Settings\textbackslash All Users\textbackslash Quick Start\textbackslash BitTorrent Sync.lnk & Application Shortcut\\ \hline
 \end{tabular}
}
\caption{BitTorrent Sync Default Application Files}
\label{tab:trace}
\end{table*}

Table~\ref{tab:hardware} shows the hardware and virtual machines used to perform an analysis on the BTSync application. The tool was installed on all machines outlined using the default installation parameters. A complete list of the files created during the install process is outlined in Table~\ref{tab:trace}.

Default installation includes the creation of a BTSync folder (the location on Windows based machines is \texttt{\$Volume\$\textbackslash Documents and Settings\textbackslash [User]\textbackslash BTSync}). This folder is automatically populated with three files:
\begin{enumerate}
\item \texttt{.SyncID} -- Stores a 20 byte unique share ID
\item \texttt{.SyncIgnore} -- A list of files in the folder or subfolder to ignore when synchronising with remote machines.
\item \texttt{.SyncArchive} (Folder) -- An archive to store files that were deleted on a remote synchronised system.
\end{enumerate}
These three files are created whenever any new BTSync share is set up and are used to aid in the control of data exchange between the nodes.

On Linux based machines, the installation directory is wherever the user chooses to unpack the application package. All of the same files are created included the hidden folders. In addition the user interface is a web GUI on \texttt{localhost:8888} and the application can generate a configuration file by running the command ``\texttt{./btsync $--$dump-sample-config}'' from the terminal. If this plain text file is edited it can be used to overwrite the username and password for the web GUI to allow the investigator access without changing any other settings.

\subsection{BTSync Client Activity}
\label{BTSyncClient}

\begin{figure}[!h]
\centering
\includegraphics[width=0.5\textwidth]{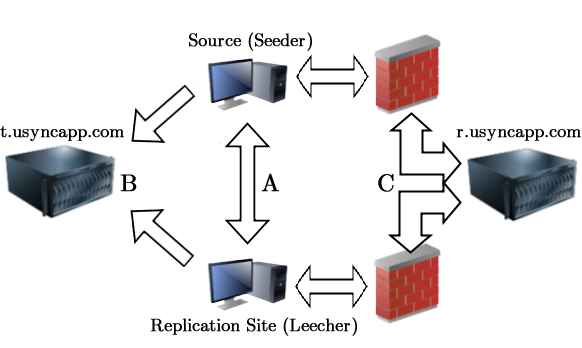}
\caption{BTSync Synchronisation Options}
\label{fig:syncopts}
\end{figure}

The options for synchronisation and replication are set for each share on the local machine. As shown in Figure~\ref{fig:syncopts}, there are three main distinct settings determining the resources used for peer discovery and the paths available for traffic transmission. BTSync uses similar peer discovery methods to the regular BitTorrent protocol. These methods are outlined below:

\begin{enumerate}
 \item LAN Discovery -- If the option \texttt{``search LAN''} is enabled the client application will start sending peer discovery packets across the LAN utilising the multicast address \texttt{IP Address: IP 239.192.0.0 Port: 3838}. These packets are sent at a frequency of one every ten seconds for each share utilising this method. 

\begin{figure}[!h]
\centering
\includegraphics[width=0.5\textwidth]{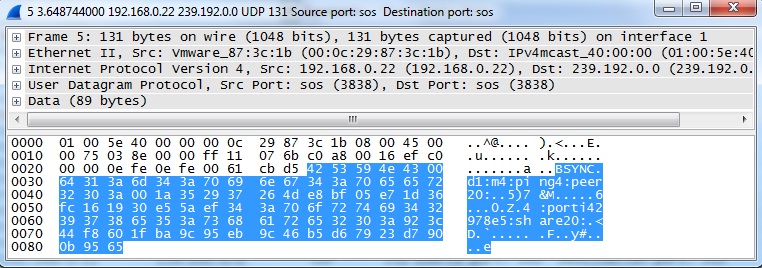}
\caption{BTSync Multicast ``Seeker'' Packet}
\label{BTSyncMCast}
\end{figure}

The local peer discovery packet has a BSYNC header and a message type of ``ping'' and includes the sending host's IP address, port and the 20 byte ShareID of the share being advertised. Hosts on the LAN receiving the packet will drop it if the ShareID is not of interest to them. Any host that has an interest will respond with a UDP packet to the port advertised. The response does not have a BSYNC header present and the data field only contains the PeerID of the responding peer. This discovery is restricted to Path `A' in Figure~\ref{fig:syncopts}.

\begin{figure}[!h]
\centering
\includegraphics[width=0.5\textwidth]{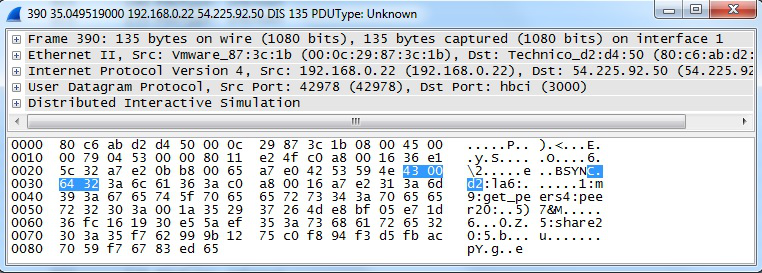}
\caption{BTSync Tracker Request Packet}
\label{BTSyncRequest}
\end{figure}

 \begin{figure}[!h]
 \centering
 \includegraphics[width=0.5\textwidth]{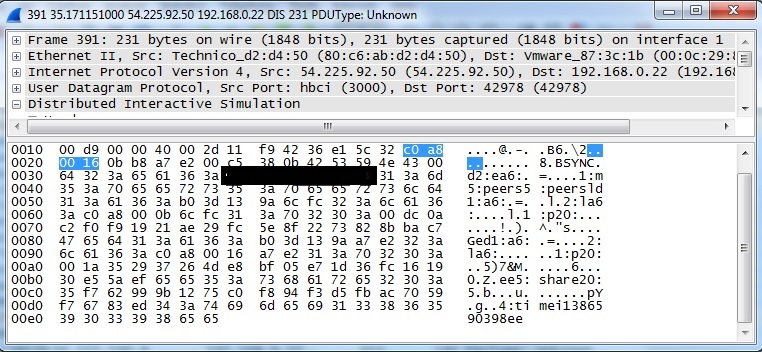}
 \caption{BTSync Tracker Response Packet}
 \label{BTSyncResponse}
 \end{figure}

\item Tracker -- The option \texttt{``Use Tracker''} causes the client to search for peers by requesting a peer list from the tracker located at \texttt{t.usyncapp.com} which was resolves to three IP addresses: 
\begin{itemize}
\item 54.225.100.8
\item 54.225.92.50
\item 54.225.196.38
\end{itemize}
These three IP addresses are each hosted on Amazon's EC2 cloud service. The client sends a \texttt{get\_peers} request to the tracker server (as can be seen in Figure~\ref{BTSyncRequest}). When this request is received, the client's IP addresses gets added to the list of active peers available for that particular ShareID on the tracker. The path to the tracker server taken by the peers is displayed as Path `B' of Figure~\ref{fig:syncopts}. The information keys contained in the \texttt{get\_peers} message are shown in Table~\ref{tab:request}. The peer discovery response, as displayed in Figure~\ref{BTSyncResponse} consists of a list of bencoded \texttt{IP:Port:PeerID:ShareID} entries identifying the known peers with the same secret. Due to the fact that the client only requests this list for a secret it already possesses, the response from the server will always contain at least one active peer, i.e., the requesting client's information.

\begin{table}[!h]
\makebox[\linewidth]{
    \begin{tabular}{|l|l|l|l|}
\hline
      \textbf{Key} & \textbf{Explanation}  \\ \hline
	d:      & [The Entire Dictionary]      \\ \hline
	la:      & [IP:Port in Network-Byte Order]      \\ \hline
	m:      & [Message Type Header, e.g., get\_peers]      \\ \hline
	peer:      & [Local Peer ID]      \\ \hline
	share:      & [Local Share ID]      \\ \hline
	e:      & [End]      \\ \hline
    \end{tabular}
}
\caption{Component Fields for Request Packet}
\label{tab:request}
\end{table}

\item Distributed Hash Table (DHT) -- The client can be set to perform peer discovery using a DHT. Using this option, any peer will register its details in the form of \\\texttt{SHA-1(Secret):IP:Port} with other peers, even those that do not participate in the swarm. Using this option a user can avoid using any form of tracking server but they may find that peer discovery is slower or less complete.
 
\item Known Peers -- The final, and least detectable, method of peer discovery is the option to \texttt{``Use Predefined Hosts''}. The user can add a list of IP address:Port combinations to the share preferences. This list of peers will be contacted directly without any lookup with a BSYNC packet containing a \texttt{ping} message type. 
\end{enumerate}

\subsection{Data Transfer}
\label{transfer}
Similar to peer discovery methods, BTSync allows the user to configure a number of options that affect how data is transferred between peers:

\begin{enumerate}
\item No options set (Path `A' in Figure~\ref{fig:syncopts}). The seeding host will attempt to communicate directly with the replication target (the leecher). This traffic is encrypted by default if it travels outside the local LAN. There is an option in the application preferences to enable or disable encryption of LAN traffic as well if the user prefers.

\item If the communication between the hosts is blocked for some reason, usually if the hosts are on different networks protected by firewalls or in segments or subnets of the same LAN locked down by inbound Access Control Lists, the option \texttt{``Use Relay Server when required''} will allow the traffic to bypass these restrictions if possible (this is represented by Path `C' in Figure~\ref{fig:syncopts}). The relay server, located at \texttt{r.usyncapp.com} resolves to:
\begin{itemize}
\item relay-01.utorrent.com (67.215.229.106)
\item relay-02.utorrent.com (67.215.231.242)
\end{itemize}

\begin{figure}[!h]
\centering
\includegraphics[width=0.5\textwidth]{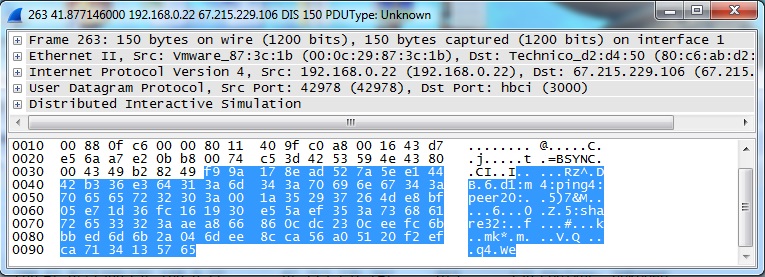}
\caption{BTSync Relay Request Packet}
\label{BTSyncRelay}
\end{figure}

\begin{figure}[!h]
\centering
\includegraphics[width=0.5\textwidth]{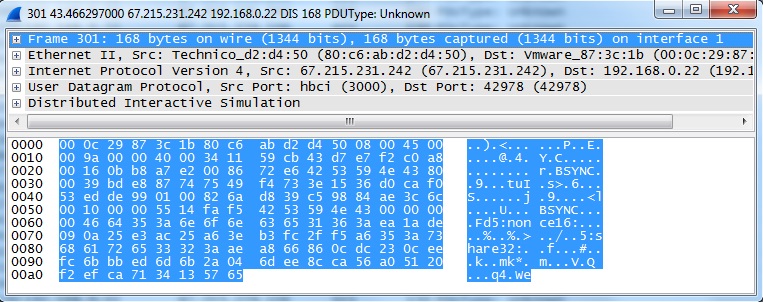}
\caption{BTSync Relay Nonce Exchange Packet}
\label{BTSyncNonce}
\end{figure}

These packets are sent via UDP to port 3000 and contain ``ping'' messages, as can be seen in Figure~\ref{BTSyncRelay} . These ping messages contain a 20 byte PeerID and a 32 byte ShareID derived from the secret key. After the initial handshake with the relay server the relay negotiates the data transmission session with the remote peer. This negotiation involves exchange of the 16 byte ``nonce'' (a one off value used for encryption purposes) and a map of the availability of the file parts, as can be seen in Figure~\ref{BTSyncNonce}. Once the handshake is complete, the next packet contains the 160 bit public key and the encrypted transfer of data begins. The responsibility for the actual data transfer is retained by the individual clients and only metadata and ping packets are sent unencrypted. 
\end{enumerate}

\subsection{BTSync Keys}
\label{BTKey}

When a share is created by a seeder, a master key is generated. This is the ``all access'', or read/write (RW), key that allows the owner of the share to add, remove or modify the contents of the share. The only scenario when this key should be distributed to another peer is when that peer is a trusted collaborator or when that peer is meant as a secondary source of content as opposed to a backup or repository. Read/write Keys can be recognised by the initial character `A' at the beginning of the 33 character ``secret'' string. All keys are stored in plaintext in the bencoded block describing the corresponding share in the \texttt{sync.dat} file. From the master key, three other types of keys can be derived:
\begin{enumerate}
\item Read Only -- The read-only (RO) key allows the receiving user to read the data being synchronised but not to modify or change the content on the source in any way. If, for some reason, a file in the share is modified or deleted on the local read-only host, its \texttt{invalidate} flag in the \texttt{<shareID>.db-wal} file is switched from a value of 0 to a value of 1. The result of this is that the file will no longer be synchronised from the source, even if the version on the source is updated or the local copy is deleted. RO keys are recognisable by the starting character `B' prepended to the 32 character secret string. It should be noted that this was originally the character `R' but it was changed with post alpha releases.
 
\item 24 Hour -- The 24 hour key can be either a RO or RW key that has a time limit of 24 hours before it expires and cannot be used. The 24 hour time limit refers to the time during which the remote peer must use the key to gain access to the share. Once used successfully the peer will have continued access until the share is deleted or the source changes the authentication key. 24 hour keys start with the character `C'. These key types are useful for security as they are only vulnerable to a third party interception for a limited period of time. The key stored in \texttt{sync.dat} is not the 24-hour key, it is the corresponding, non-expiring RW or RO equivalent.
 
 \item Encrypted -- There is an encrypted key that can be generated that creates an encrypted repository on the remote peer. The files synchronised are stored in their encrypted state remotely and cannot be read by the operator of the remote machine unless they are given the decryption key as well. This type of key is only possible to produce if the developer API has been installed and further analysis is outside the scope of this paper. Investigators should be aware that an encryption key is recognisable by the character `D' at the start of the 33 character sequence.
\end{enumerate}
 
In addition to the key strings, BTSync gives users the option of creating a RW or RO QR code that they can scan into a mobile application if preferred. Seeders must be very careful about what keys they distribute and to whom they are distributed. A RW key sent to the wrong person could subvert the assurance of file integrity and negate many of the benefits of BTSync over a shared folder hosted at a neutral location. 

Sample keys taken from the same BTSync Share:\\
RW: \texttt{ACHY3VFJZ3RJ3DE2CHPUGE6W7EZSRA3OR}\\
RO: \texttt{BY6G6B7KIBGELLXE2RL65C34CAGPV7LUJ}\\
24-hour RW: \texttt{CBJIK32CLMWF2P7JLFYRGC3JRTEZ6JLPU}\\
24-hour RO: \texttt{CCYGZN6R67O67QB7HGLL4F5BAVA3AJ5LC}

%%%%%%%%%%%%%%%%%%%%%%%%%%%%%%%%%%%%%%%%%%%%%%%%%%%%%%%%%%%%%%%
\section{Sources of Interest to Forensic Investigation}
\label{sources}
%%%%%%%%%%%%%%%%%%%%%%%%%%%%%%%%%%%%%%%%%%%%%%%%%%%%%%%%%%%%%%%
To determine what can be found without resorting to specialist forensic utilities the BTSync application was installed and three folders were synchronised. The default settings were chosen at installation which include:
\begin{itemize}
\item BTSync runs at startup.
\item BTSync service icon in the system tray (right click to hide).
\item BTSync shortcut placed on the desktop of the All Users profile.
\item BTSync added to the ``All Users'' profile quick launch.
\end{itemize}

In order to gather sample network data, three separate synchronisations were set up and monitored:
\begin{enumerate}
\item To \texttt{\$Volume\$\textbackslash Documents and Settings\textbackslash [User] \textbackslash Desktop\textbackslash sharedfolder} from a separate Linux laptop on the same LAN.
\item From \texttt{\$Volume\$\textbackslash Documents and Settings\textbackslash [User] \textbackslash Desktop\textbackslash sf2} on localhost to a separate Linux laptop on the same LAN.
\item Performed using a secret key posted on Reddit \cite{reddit}. The folder advertised itself as containing Gameboy ROMs with the read-only shared key of ``RUAM2ED5ISKYR7LVELNVX56LLHQ47GBOZ''. The application does not provide an indication as to what size the remote folder is or what files it contains before commencing the download.
\end{enumerate}

As each folder was shared and assigned a secret key (either generated locally or copied from another source) a file was created in the folder:
\texttt{\$Volume\$\textbackslash Documents and Settings\textbackslash [User]\textbackslash Application Data\textbackslash BitTorrent Sync\textbackslash} with the ShareID of the folder created. This is the same share ID found in the file \texttt{.SyncID} created in the share folder itself.

The name of the db files created when the shared folder was added to BTSync consisted of the contents of the \texttt{.SyncID} file (35F762999B1275C0F894F3D5FBAC7059F76783ED). This is the 20 byte share code that gets advertised to \texttt{t.usyncapp.com} when BTSync sends out its \texttt{get\_peer} message, as can be seen in Figure~\ref{BTSyncRequest}. 

As each synchronisation was run, the BTSync logfile located at \texttt{\$Volume\$\textbackslash Documents and Settings\textbackslash [User]\textbackslash Application Data\textbackslash Bittorrent Sync\textbackslash sync.log} is updated to record events. A sample of what this log filed contains is outlined in Table~\ref{tab:synclog}. The behaviour seen in the sync.log file corresponds with that observed in the captured network activity and the system activity recorded.
\begin{table}[t]
\begin{tabular}{|>{\raggedright\arraybackslash}p{8.5cm}|}
\hline
[2013-12-01 12:41:33] Loading config file version 1.1.82
\\ \hline
[2013-12-01 12:41:33] Loaded folder \textbackslash \textbackslash ?\textbackslash {\fontfamily{ptm}\selectfont\texttildelow}User\textbackslash BTSync
\\ \hline
[2013-12-01 12:41:33] Loaded folder \textbackslash \textbackslash ?\textbackslash {\fontfamily{ptm}\selectfont\texttildelow}User\textbackslash Desktop\textbackslash sharefolder
\\ \hline
[2013-12-01 12:41:33] Loaded folder \textbackslash \textbackslash ?\textbackslash {\fontfamily{ptm}\selectfont\texttildelow}User\textbackslash Desktop\textbackslash sf2
\\ \hline
[2013-12-01 12:43:44] Got ping (broadcast: 1) from peer 192.168.0.11:27900 (00DC0AC2F0F91921AE29FC5E8F2273828BBAC747) for share 35F762999B1275C0F894F3D5FBAC7059F76783ED
\\ \hline
[2013-12-01 12:43:44] Found peer for folder \textbackslash \textbackslash ?\textbackslash {\fontfamily{ptm}\selectfont\texttildelow}User\textbackslash Desktop\textbackslash sharefolder 00DC0AC2F0F91921AE29FC5E8F2273828BBAC747 192.168.0.11:27900 direct:1
\\ \hline
[2013-12-01 12:43:45] Sending broadcast ping for share 55045F90CA4C1A42DDB78DCD132F3ACC33E946EC
\\ \hline
[2013-12-01 12:43:45] Requesting peers from server
\\ \hline
[2013-12-01 12:43:45] Sending broadcast ping for share 35F762999B1275C0F894F3D5FBAC7059F76783ED
\\ \hline
\end{tabular}
    \caption {Sample Contents of BitSync Log File}
\label{tab:synclog}
\end{table}

Table~\ref{tab:synchro} presents the system activity logged during the synchronisation process. This was performed in a monitored session whereby a text file named ``sample3.txt'' was created on the source host (seeder) and then the read/write secret was shared to the prepared receiving folder on the repository (leecher).
The synchronization process is shown from the point where apply was clicked on the repository. In the table \texttt{AppData} is shorthand for \\\texttt{{\fontfamily{ptm}\selectfont\texttildelow}User\textbackslash Application Data\textbackslash Bittorrent Sync} and \texttt{Share} represents the path to the folder that has been allocated to receive the data. In this particular instance it is \texttt{C$:$\textbackslash Documents and Settings\textbackslash User\textbackslash Desktop\textbackslash sharedfolder}.

The shared folder is populated with the application control files and the 20 byte shareID is written to the \texttt{.SyncID} file.The database files are created in the User application data folder. The filenames used for these database files are the same as the ShareID stored in the \texttt{.SyncID} file. \texttt{.SyncIgnore} is created in the share folder and 822bytes are written to it. The data written are the explanation of the file's purpose and usage as well as a short list of
files usually generated by an Operating System.

Next the synchronization process starts with the creation of \texttt{sync.dat.new} which will be renamed to \texttt{sync.dat} and eventually \texttt{sync.dat.old} as subsequent synchronisations take place. The \texttt{<ShareID>.db-wal} file is created to act as a holding area for data to be written to the SQLite database file of the same name. Next the data is received and written to a synchronisation delta file in preparation for merging into a fully synchronized text file. File data waiting merger can be identified by the extension \texttt{$!$sync} and \texttt{$!$sync(X)}.

\begin{table}[t]
\begin{tabular}{|l|p{1.5in}|l|l|}
\hline
\textbf{Action} & \textbf{File}  & \textbf{I/O}  &  \textbf{Path}  \\ \hline
Create & \texttt{\small .SyncID}  & 20B  & \texttt{Share} \\ \hline
Create & \texttt{\small <ShareID>.db}  & \- & \texttt{AppData} \\ \hline
Create & \texttt{\small <ShareID>.db-journal} & \- & \texttt{AppData} \\ \hline
Write &   \texttt{\small <ShareID>.db-journal} & 512B & \texttt{AppData} \\ \hline
Write &  \texttt{\small <ShareID>.db} & 3 KB & \texttt{AppData} \\ \hline
Delete & \texttt{\small <ShareID>.db-journal} & \- & \texttt{AppData} \\ \hline
Create & \texttt{\small .SyncIgnore} & 822B & \texttt{Share} \\ \hline
Create & \texttt{\small sync.dat.new} & 822B & \texttt{AppData} \\ \hline
Rename & \texttt{\small sync.dat to} \texttt{sync.dat.old} & 450B & \texttt{AppData} \\ \hline
Rename & \texttt{\small sync.dat.new to} \texttt{sync.dat} & 822B & \texttt{AppData} \\ \hline
Create & \texttt{\small <ShareID>.db-wal} & \- & \texttt{AppData} \\ \hline
Create & \texttt{\small sample3.txt.$!sync$} & 33B & \texttt{Share} \\ \hline
Rename & \texttt{\small sample3.txt.$!sync$ to} \texttt{sample3.txt.$!sync.!sync1$} & 33B & \texttt{Share} \\ \hline
Write & \texttt{\small sample3.txt.\!sync.\!sync1} & 33B & \texttt{Share} \\ \hline
Rename & \texttt{\small sample3.txt.\!sync.\!sync1} to:\texttt{sample3.txt} & \- & \texttt{Share} \\ \hline
\end{tabular}
\caption{Example File I/O During the Client's Synchronisation Procedure}
\label{tab:synchro}
\end{table}

 The registry keys outlined in Table~\ref{tab:regkeys} were found after installation.
\begin{table}[h]
\begin{tabular}{|>{\raggedright\arraybackslash}p{8.5cm}|}

\hline
    HKCR $\backslash$Applications  $\backslash$BTSync.exe $\backslash$shell $\backslash$open $\backslash$command                         \\ \hline
    HKCU $\backslash$Software $\backslash$Classes $\backslash$Applications $\backslash$BTSync.exe $\backslash$shell $\backslash$open $\backslash$command \\ \hline
    HKCU $\backslash$Software $\backslash$Microsoft $\backslash$Windows $\backslash$CurrentVersion $\backslash$Run                      \\ \hline
    HKCU $\backslash$Software $\backslash$Microsoft $\backslash$Windows $\backslash$ShellNoRoam $\backslash$MUICache                    \\ \hline
    HKLM $\backslash$SOFTWARE $\backslash$Microsoft $\backslash$ESENT $\backslash$Process $\backslash$BTSync $\backslash$DEBUG <--if debug log enabled \\ \hline
    HKLM $\backslash$SOFTWARE $\backslash$Microsoft $\backslash$Windows $\backslash$CurrentVersion $\backslash$Uninstall $\backslash$BitTorrent Sync \\ \hline
    HKLM $\backslash$SYSTEM $\backslash$ControlSet001 $\backslash$Services $\backslash$SharedAccess $\backslash$Parameters $\backslash$FirewallPolicy $\backslash$StandardProfile $\backslash$AuthorizedApplications $\backslash$List \\ ~ \\value: (C: $\backslash$Program Files $\backslash$BitTorrent Sync $\backslash$BTSync.exe:*:Enabled:BitTorrent Sync)                           \\ \hline
    HKU $\backslash$S-1-5-21...-1003 $\backslash$Software $\backslash$Classes $\backslash$Applications $\backslash$BTSync.exe \\ \hline
    HKU $\backslash$S-1-5-21...-1003 $\backslash$Software $\backslash$Classes $\backslash$Applications $\backslash$BTSync.exe $\backslash$shell $\backslash$open $\backslash$command \\ \hline
    HKU $\backslash$S-1-5-21...-1003 $\backslash$Software $\backslash$Microsoft $\backslash$Windows $\backslash$CurrentVersion $\backslash$Run \\ \hline
    HKU $\backslash$S-1-5-21...-1003 $\backslash$Software $\backslash$Microsoft $\backslash$Windows $\backslash$ShellNoRoam $\backslash$MUICache \\ \hline
    HKU $\backslash$S-1-5-21...-1003\_Classes $\backslash$Applications $\backslash$BTSync.exe $\backslash$shell $\backslash$open $\backslash$command \\ \hline
    C: $\backslash$Program Files $\backslash$BitTorrent Sync $\backslash$BTSync.exe                                                              \\ \hline
    C: $\backslash$Documents and Settings $\backslash$All Users $\backslash$Desktop $\backslash$BTSync.lnk                                          \\ \hline
    \end{tabular}
    \caption {Created BTSync Registry Keys During Installation}
\label{tab:regkeys}
\end{table}

Next a file was deleted from the remote host and ten minutes were given to ensure the local host had synchronised completely. While the file had been removed completely from the original host, on the local host it was instead moved from the main folder to a hidden subfolder (\texttt{.SyncArchive}) and not moved to the recycle bin. It is unknown at this time if there is any trigger or flag set that would result in this hidden file being deleted completely off the system. If not, then a remote host could theoretically constantly add and remove files to a synchronisation folder in order to deliberately occupy space on the local host with hidden files and so perform a form of low-tech denial of service attack by filling local storage.

BTSync does not come with any uninstaller of its own and must be removed from the Control panel. After uninstall the system was rebooted to ensure that the service had stopped running and any post uninstall clean-up had been performed, file locks cleared etc. A number of associated registry keys were still present, as outlined in Table~\ref{tab:regkeysafteruninstall}.

\begin{table}[h]
\begin{tabular}{|>{\raggedright\arraybackslash}p{8.5cm}|}
\hline
HKCR $\backslash$Applications $\backslash$BTSync.exe $\backslash$shell $\backslash$open $\backslash$command
\\ \hline
HKCU $\backslash$Software $\backslash$Classes $\backslash$Applications $\backslash$BTSync.exe $\backslash$shell $\backslash$open $\backslash$command
\\ \hline
HKCU $\backslash$Software $\backslash$Microsoft $\backslash$Windows $\backslash$CurrentVersion $\backslash$Run
\\ \hline
("C: $\backslash$Program Files $\backslash$BitTorrent Sync $\backslash$BTSync.exe" /MINIMIZED)
\\ \hline
HKCU $\backslash$Software $\backslash$Microsoft $\backslash$Windows $\backslash$ShellNoRoam $\backslash$MUICache
\\ \hline
HKLM $\backslash$SOFTWARE $\backslash$Microsoft $\backslash$ESENT $\backslash$Process $\backslash$BTSync $\backslash$DEBUG 
\\ \hline
(BTSync Rot 13 encoded = OGflap)
\\ \hline
HKCU $\backslash$Software $\backslash$Microsoft $\backslash$Windows $\backslash$CurrentVersion $\backslash$Explorer $\backslash$UserAssist $\backslash${75048700-EF1F-11D0-9888-006097DEACF9} $\backslash$Count      Key =  HRZR\_EHACNGU:P: $\backslash$Qbphzragf naq Frggvatf $\backslash$BFv $\backslash$Qrfxgbc $\backslash$OGFlap.rkr
\\ \hline
HKU $\backslash$S-1-5-21...-1003 $\backslash$Software $\backslash$Microsoft $\backslash$Windows $\backslash$CurrentVersion $\backslash$Explorer $\backslash$UserAssist $\backslash${75048700-EF1F-11D0-9888-006097DEACF9} $\backslash$Count      Key =  HRZR\_EHACNGU:P: $\backslash$Qbphzragf naq Frggvatf $\backslash$BFv $\backslash$Qrfxgbc $\backslash$OGFlap.rkr
\\ \hline
    \end{tabular}
    \caption {Registry Keys Remaining After Uninstallation}
\label{tab:regkeysafteruninstall}
\end{table}

In addition to this, all shared file folders used in synchronisations were still present as well as the default BTSync share created at install. The application folder was also still present in the \texttt{\$Volume\$\textbackslash Documents and Settings\textbackslash [User]\textbackslash Application Data} folder but the sync.log file had been emptied.

As well as registry keys BTSync creates several other files that may be of interest to the forensic investigator. These files are located in the directory \texttt{\$Volume\$\textbackslash Documents and Settings\textbackslash [User]\textbackslash Application Data\textbackslash Bittorrent Sync\textbackslash}. The contents of each file is outlined below:

\begin{itemize}
\item \texttt{<40 character share ID number>[.db, .db-shm, .db-wal]} -- These files contribute to a SQLite3 database. The database describes the contents of the share directory corresponding to the share ID. It contains filenames, transfer piece registers and hash values for each individual file and its constituent pieces. While the .db file stores information on the schema of the database the db-wal file contains bencoded entries for each file within the share in the format:

\texttt{<Filename>:invalidated1:main} \\
\texttt{hash:<20 byte hash>:mtime:} \\
\texttt{<timestamp of modification time>:npieces1:} \\
\texttt{owner20:<20 byte PeerID of the Seeder>:} \\
\texttt{path<path to file within share>} \\
\texttt{perm:420:size[bytes]:state1:timestamp:type1}  \\
\texttt{pvtime0:sig:<32 byte signature><filename>} \\

\item \texttt{settings.dat} -- This is a bencoded file with a fileguard key (this key is a salted hash value ensuring that the file has not been edited by another tool besides the BTSync application itself). This file contains a log of settings for the application including the settings used to generate the Cryptographic keys and the registered URLs for peer searches.
\item \texttt{sync.dat} -- This is a bencoded file with a fileguard key. This file lists what files have been synchronised across the network. The directory paths and the shared secret used can be recovered from this file. This file is perhaps of most interest to the investigator due ot the large amount of timestamped and option recording it contains. Each share has an entry that is laid out in the following style:

\texttt{path:<full path to share folder>:} \\
\texttt{secret:<33 character Key>:} \\
\texttt{pub\_key:<32 byte ShareID used in Relay messages>:} \\
\texttt{stopped\_by\_user[0|1]:} \\
\texttt{use\_dht[0|1]:use\_lan\_broadcast[0|1]:} \\
\texttt{use\_relay[0|1]:use\_tracker[0|1]:} \\
\texttt{use\_known\_hosts[0|1]:} \\
\texttt{known\_hosts:<contents of known hosts option>:} \\
\texttt{peers:<list of peerIDs involved in sync>:} \\
\texttt{last\_sync\_completed<timestamp>:} \\
\texttt{invites<list of swarm invites received>:} \\
\texttt{folder\_type0:} \\
\texttt{delete\_to\_trash[0|1]:} \\
\texttt{mutex\_file\_initialized[0|1]:} \\
\texttt{directTotal<IO direct to/from peer>:} \\
\texttt{relayTotal<IO total between peer and relay>} \\

\item \texttt{settings.dat.old} -- This is the previous settings file. BTSync rotates through two settings generations deleting the old file when it is time to update with new data.
\end{itemize}

\subsection{Recovering Destroyed Evidence}
\label{deb}

A number of the above artefacts prove that BTSync was installed on a client machine. It is possible that some or all of the incriminating files themselves may prove unrecoverable from the local hard disk due to anti-forensic measures. Should the secret be recovered for a given share, it is possible that a synchronisation of the suspect secret will enable the forensic investigator to recover this lost information from any other nodes still active in the share. Regular file system forensic analysis identifying synchronisation artefacts left behind from a particular share combined with this subsequent data synchronisation can prove that the suspect machine was involved in the sharing of incriminating material. Like any other digital investigation, the evidence gathered from the synchronisation process should be collected into a suitable digital evidence bag. Due to the value of BTSync metadata in the recovery of files stored remotely, a suitable P2P oriented evidence bag should be selected, such as that proposed by Scanlon and Kechadi \cite{scanlon2014digital}. The after-the-fact recovery of data from remote machines is beyond the scope of this paper.

%%%%%%%%%%%%%%%%%%%%%%%%%%%%%%%%%%%%%%%%%%%%%%%%%%%%%%%%%%%%%%%
\section{Conclusion}
\label{conclusion}
%%%%%%%%%%%%%%%%%%%%%%%%%%%%%%%%%%%%%%%%%%%%%%%%%%%%%%%%%%%%%%%

This paper gave a first look at a new use for a popular and widespread file synchronisation protocol. BTSync is not intended to replace BitTorrent as a file dissemination utility. However, it is still being used for this purpose. This is facilitated though websites publicly providing shared secrets, e.g., Reddit \cite{reddit}, as a form of dead-drop. The developers of the application describe it as an end-to-end encrypted method of transferring files without the use of a third party staging area. The reason for this is to try and ensure that the content and personal details remain hidden from unauthorised access. Initial analysis of the installation procedure identified files most likely to be of use to a forensic examiner confronted with a suspect live system or image running BTSync. However while the presence of a SyncID hidden folder can explain how a file was transferred to a system there is currently no way known outside of the application itself to determine the file's origin or any further synchronisation points. From an investigative perspective, the decentralised nature of BTSync will always leave an avenue of gathering information identifying nodes sharing particular content (while providing many desirable redundancy and resilience against attack).

For the digital investigator working on a case involving BTSync, the description of the registry keys and files outlined can aid in identifying the content that may have been present on the local machine. Seeing as though BTSync merely requires any user wishing to join a particular synchronised folder to have the key, an investigator could also join the shared folder to download the data having recovered the corresponding files through hard drive analysis. Subsequent monitoring of the network communications using common tools, e.g., WireShark, tcpdump or libpcap, can aid in the identification of other nodes syncing the same content. In a number of investigative scenarios, this may focus the investigation in a beneficial direction resulting in the discovery of additional pertinent evidence or additional suspects.

\subsection{Future Work}
\label{futurework}
From this initial analysis of the BTSync system, much further work needs to be done. The following list amounts to the list of areas for future investigation:

\begin{itemize}
\item Network Analysis -- Identification of BTSync traffic and subsequent analysis to determine differentiation from standard BitTorrent traffic.
\item Investigation Utility -- A standalone application to extract relevant information from a suspect live or imaged machine running BTSync.
\item Automated Share Detection -- Identifying BTSync shares advertised by BTSync clients and detecting network activity to or from known locations.
\item Crawling -- To systematically follow connections to or from a share and identify new connections as they are discovered.
\item Enumeration -- Identifying individual shares and all active swarm members by the participating IP addresses and peer identifiers.
\item Geolocation -- Geolocating identified IP addresses may prove pertinent to recovering additional evidence regarding the suspect or may aid in the identification of others involved.
\item API Analysis -- Testing the provisioned API to determine what features can be leveraged to assist in forensic investigations.
\item Recovery of Deleted Shares -- In the scenario where a suspect has securely deleted any incriminating evidence from the local machine, the identification of trace information on the machine may result in the evidence being recoverable from other remote hosts. Due to BitTorrent's reliance on regular hashing for file distribution, the resultant hashes of remotely gathered files may be resolvable to the suspect's machine.
\end{itemize}

\bibliographystyle{model6-num-names}
\bibliography{bibfile}

\begin{thebibliography}{16}
\providecommand{\natexlab}[1]{#1}
\providecommand{\url}[1]{\texttt{#1}}
\providecommand{\href}[2]{#2}
\providecommand{\path}[1]{#1}
\providecommand{\DOIprefix}{doi:}
\providecommand{\ArXivprefix}{arXiv:}
\providecommand{\URLprefix}{URL: }
\providecommand{\Pubmedprefix}{pmid:}
\providecommand{\doi}[1]{\href{http://dx.doi.org/#1}{\path{#1}}}
\providecommand{\Pubmed}[1]{\href{pmid:#1}{\path{#1}}}
\providecommand{\BIBand}{and}
\providecommand{\bibinfo}[2]{#2}
\ifx\xfnm\undefined \def\xfnm[#1]{\unskip,\space#1}\fi
\makeatletter\def\@biblabel#1{#1.}\makeatother
%Type = Misc
\bibitem[{{BitTorrent Inc}(2013{\natexlab{a}})}]{bitsync}
\bibinfo{author}{{BitTorrent Inc}\xfnm[]}.
\newblock \bibinfo{title}{{BitTorrent Sync User Manual}}.
\newblock
  \bibinfo{howpublished}{\\\url{http://www.bittorrent.com/help/manual/}};
  \bibinfo{year}{2013}{\natexlab{a}}.
\newblock \bibinfo{note}{[Online; accessed February 2014]}.
%Type = Misc
\bibitem[{{BitTorrent Inc}(2013{\natexlab{b}})}]{bitsyncapi}
\bibinfo{author}{{BitTorrent Inc}\xfnm[]}.
\newblock \bibinfo{title}{{BitTorrent Sync Developer API}}.
\newblock
  \bibinfo{howpublished}{\\\url{http://www.bittorrent.com/sync/developers/api}};
  \bibinfo{year}{2013}{\natexlab{b}}.
\newblock \bibinfo{note}{[Online; accessed February 2014]}.
%Type = Misc
\bibitem[{{BitTorrent Inc}(2013{\natexlab{c}})}]{bitsyncstats}
\bibinfo{author}{{BitTorrent Inc}\xfnm[]}.
\newblock \bibinfo{title}{{BitTorrent Sync Article}}.
\newblock
  \bibinfo{howpublished}{\\\url{http://blog.bittorrent.com/2013/12/05/bittorrent-sync-hits-2-million-user-mark/}};
  \bibinfo{year}{2013}{\natexlab{c}}.
\newblock \bibinfo{note}{[Online; accessed February 2014]}.
%Type = Misc
\bibitem[{Cohen(2014)}]{cohen2008bittorrent}
\bibinfo{author}{Cohen\xfnm[ B.]}.
\newblock \bibinfo{title}{{The BitTorrent Protocol Specification and
  Enhancement Proposals}}.
\newblock
  \bibinfo{howpublished}{\\\url{http://www.bittorrent.org/beps/bep_0000.html}};
  \bibinfo{year}{2014}.
\newblock \bibinfo{note}{[Online; accessed February 2014]}.
%Type = Inproceedings
\bibitem[{Stutzbach and Rejaie(2006)}]{Stutzbach:2006:UCP:1177080.1177105}
\bibinfo{author}{Stutzbach\xfnm[ D.]}, \bibinfo{author}{Rejaie\xfnm[ R.]}.
\newblock \bibinfo{title}{Understanding churn in peer-to-peer networks}.
\newblock In: \emph{\bibinfo{booktitle}{Proceedings of the 6th ACM SIGCOMM
  Conference on Internet Measurement}}. IMC '06; \bibinfo{address}{New York,
  NY, USA}: \bibinfo{publisher}{ACM}.
\newblock ISBN \bibinfo{isbn}{1-59593-561-4}; \bibinfo{year}{2006}:\unskip
  \bibinfo{pages}{189--202}.
\newblock \URLprefix \url{http://doi.acm.org/10.1145/1177080.1177105}.
  \DOIprefix\doi{10.1145/1177080.1177105}.
%Type = Article
\bibitem[{Karakaya et~al.(2009)Karakaya, Korpeoglu and Ulusoy}]{4797943}
\bibinfo{author}{Karakaya\xfnm[ M.]}, \bibinfo{author}{Korpeoglu\xfnm[ I.]},
  \bibinfo{author}{Ulusoy\xfnm[ O.]}.
\newblock \bibinfo{title}{Free riding in peer-to-peer networks}.
\newblock \emph{\bibinfo{journal}{Internet Computing, IEEE}}
  \bibinfo{year}{2009};\bibinfo{volume}{13}(\bibinfo{number}{2}):\bibinfo{pages}{92--98}.
\newblock \DOIprefix\doi{10.1109/MIC.2009.33}.
%Type = Article
\bibitem[{Scanlon et~al.(2010)Scanlon, Hannaway and Kechadi}]{scanlon2010week}
\bibinfo{author}{Scanlon\xfnm[ M.]}, \bibinfo{author}{Hannaway\xfnm[ A.]},
  \bibinfo{author}{Kechadi\xfnm[ M.T.]}.
\newblock \bibinfo{title}{{A Week in the Life of the Most Popular BitTorrent
  Swarms}}.
\newblock \emph{\bibinfo{journal}{5th Annual Symposium on Information Assurance
  (ASIA'10)}} \bibinfo{year}{2010};.
%Type = Inproceedings
\bibitem[{Cohen(2003)}]{cohen2003incentives}
\bibinfo{author}{Cohen\xfnm[ B.]}.
\newblock \bibinfo{title}{Incentives build robustness in bittorrent}.
\newblock In: \emph{\bibinfo{booktitle}{Proceedings of the Workshop on
  Economics of Peer-to-Peer systems}}; vol.~\bibinfo{volume}{6}.
  \bibinfo{year}{2003}:\unskip \bibinfo{pages}{68--72}.
%Type = Article
\bibitem[{Chung et~al.(2012)Chung, Park, Lee and Kang}]{Chung201281}
\bibinfo{author}{Chung\xfnm[ H.]}, \bibinfo{author}{Park\xfnm[ J.]},
  \bibinfo{author}{Lee\xfnm[ S.]}, \bibinfo{author}{Kang\xfnm[ C.]}.
\newblock \bibinfo{title}{Digital forensic investigation of cloud storage
  services}.
\newblock \emph{\bibinfo{journal}{Digital Investigation}}
  \bibinfo{year}{2012};\bibinfo{volume}{9}(\bibinfo{number}{2}):\bibinfo{pages}{81
  -- 95}.
%Type = Article
\bibitem[{Quick and Choo(2013{\natexlab{a}})}]{Quick2013266}
\bibinfo{author}{Quick\xfnm[ D.]}, \bibinfo{author}{Choo\xfnm[ K.K.R.]}.
\newblock \bibinfo{title}{Forensic collection of cloud storage data: Does the
  act of collection result in changes to the data or its metadata?}
\newblock \emph{\bibinfo{journal}{Digital Investigation}}
  \bibinfo{year}{2013}{\natexlab{a}};\bibinfo{volume}{10}(\bibinfo{number}{3}):\bibinfo{pages}{266
  -- 277}.
%Type = Article
\bibitem[{Quick and Choo(2013{\natexlab{b}})}]{Quick2013}
\bibinfo{author}{Quick\xfnm[ D.]}, \bibinfo{author}{Choo\xfnm[ K.K.R.]}.
\newblock \bibinfo{title}{Google drive: Forensic analysis of data remnants}.
\newblock \emph{\bibinfo{journal}{Journal of Network and Computer
  Applications}} \bibinfo{year}{2013}{\natexlab{b}};.
%Type = Article
\bibitem[{Quick and Choo(2013{\natexlab{c}})}]{quick2013digital}
\bibinfo{author}{Quick\xfnm[ D.]}, \bibinfo{author}{Choo\xfnm[ K.K.R.]}.
\newblock \bibinfo{title}{{Digital Droplets: Microsoft SkyDrive Forensic Data
  Remnants}}.
\newblock \emph{\bibinfo{journal}{Future Generation Computer Systems}}
  \bibinfo{year}{2013}{\natexlab{c}};.
%Type = Inproceedings
\bibitem[{Drago et~al.(2012)Drago, Mellia, M.~Munafo, Sperotto, Sadre and
  Pras}]{Drago:2012:IDU:2398776.2398827}
\bibinfo{author}{Drago\xfnm[ I.]}, \bibinfo{author}{Mellia\xfnm[ M.]},
  \bibinfo{author}{M.~Munafo\xfnm[ M.]}, \bibinfo{author}{Sperotto\xfnm[ A.]},
  \bibinfo{author}{Sadre\xfnm[ R.]}, \bibinfo{author}{Pras\xfnm[ A.]}.
\newblock \bibinfo{title}{Inside dropbox: Understanding personal cloud storage
  services}.
\newblock In: \emph{\bibinfo{booktitle}{Proceedings of the 2012 ACM Conference
  on Internet Measurement Conference}}. IMC '12; \bibinfo{address}{New York,
  NY, USA}: \bibinfo{publisher}{ACM}.
\newblock ISBN \bibinfo{isbn}{978-1-4503-1705-4}; \bibinfo{year}{2012}:\unskip
  \bibinfo{pages}{481--494}.
%Type = Inproceedings
\bibitem[{Wang et~al.(2012)Wang, Shea, Wang and Liu}]{wang2012impact}
\bibinfo{author}{Wang\xfnm[ H.]}, \bibinfo{author}{Shea\xfnm[ R.]},
  \bibinfo{author}{Wang\xfnm[ F.]}, \bibinfo{author}{Liu\xfnm[ J.]}.
\newblock \bibinfo{title}{On the impact of virtualization on dropbox-like cloud
  file storage/synchronization services}.
\newblock In: \emph{\bibinfo{booktitle}{Proceedings of the 2012 IEEE 20th
  International Workshop on Quality of Service}}. \bibinfo{organization}{IEEE
  Press}; \bibinfo{year}{2012}:\unskip~\bibinfo{pages}{11}.
%Type = Misc
\bibitem[{{Reddit}(2013)}]{reddit}
\bibinfo{author}{{Reddit}\xfnm[]}.
\newblock \bibinfo{title}{{BTSecrets}}.
\newblock \bibinfo{howpublished}{\\\url{http://www.reddit.com/r/btsecrets}};
  \bibinfo{year}{2013}.
\newblock \bibinfo{note}{[Online; accessed February 2014]}.
%Type = Incollection
\bibitem[{Scanlon and Kechadi(2014)}]{scanlon2014digital}
\bibinfo{author}{Scanlon\xfnm[ M.]}, \bibinfo{author}{Kechadi\xfnm[ M.T.]}.
\newblock \bibinfo{title}{{Digital Evidence Bag Selection for P2P Network
  Investigation}}.
\newblock In: \emph{\bibinfo{booktitle}{Future Information Technology}}.
  \bibinfo{publisher}{Springer}; \bibinfo{year}{2014}:\unskip
  \bibinfo{pages}{307--314}.

\end{thebibliography}

\end{document}